\documentclass[pss,fleqn,embeddedheads]{w-art}
\usepackage{times}
\usepackage{w-thm}
\usepackage[]{graphicx}
\begin{document}
\DOIsuffix{theDOIsuffix}
 \Volume{XX}
 \Issue{X}
 \Month{XX}
 \Year{2006} 
\pagespan{1}{} 
\Receiveddate{***} \Reviseddate{***} \Accepteddate{***}
\Dateposted{***} 
\subjclass[pacs]{78.67.Hc, 42.65.Re, 71.38.-k, 71.35.-y}

\title[Nonlinear optical response and exciton
dephasing in quantum dots]{Nonlinear optical response and exciton
dephasing \\  in quantum dots}
\author[E. A. Muljarov]{E. A. Muljarov\footnote{Corresponding
     author: e-mail: {\sf muljarov@gpi.ru}, Phone: +49\,30\,2093\,4992,
     Fax: +49\,30\,2093\,4725}\inst{1,2}}
     \address[\inst{1}]{Institut f\"ur Physik der Humboldt-Universit\"at zu
Berlin, Newtonstra\ss e 15, 12489 Berlin, Germany} 
\author[R. Zimmermann]{R. Zimmermann\inst{1}}
     \address[\inst{2}]{A.M.\,Prokhorov General Physics Institute RAS,
Vavilova 38, Moscow 119991, Russia } 
\begin{abstract}
The full time-dependent four-wave mixing polarization in quantum
dots is microscopically calculated, taking into account acoustic
phonon-assisted transitions between different exciton states of
the dot. It is shown that quite different dephasing times of
higher exciton states in pancake anisotropic InGaAs quantum dots
are responsible for the experimentally observed~\cite{Borri05OECS}
double-exponential decay in the photon echo signal.
\end{abstract}
\maketitle                   

\renewcommand{\leftmark}
{E. A. Muljarov and R. Zimmermmann: Exciton nonlinear response in
quantum dots}

\section{Introduction}

Dephasing of the optical polarization caused by the interaction
with lattice vibrations (phonons) is inevitable in solid state
nanostructures and presents a fundamental obstacle for their
application in quantum computing. Exciton dephasing in InGaAs
quantum dots (QDs) has been recently studied  by time-integrated
four-wave mixing (FWM) measurements~\cite{Borri01}. For the ground
state excitonic transition in QDs, the temporal dynamics of the
measured nonlinear polarization
shows an initial rapid decay within the first few picoseconds
which is followed at later times by a much slower exponential
decay of the zero-phonon line~\cite{Borri01, Borri05}. Excited
states, in turn, have a quite different behavior. It has been
recently shown~\cite{Borri05OECS} that after an initial fast decay
(similar to that of the ground state) the excited state
polarization has a double-exponential decay. A related
non-exponential long-time decay has been also observed in QD
molecules~\cite{Borri03}.

The initial decay has been studied theoretically by Vagov et
al.~\cite{Vagov02, Vagov03}, taking into account the interaction
of a single exciton state of a QD with bulk acoustic phonons, and
using the exactly solvable independent boson model
(BM)~\cite{Mahan}. These results reproduce well the experimental
FWM signal at short delay times~\cite{Vagov04}. However, there is
no long-time decay in the BM.  In order to go beyond one has to
consider the phonon-assisted coupling between different excitonic
states in the QD. Recently we have developed a microscopic
approach for the linear excitonic polarization in QDs, taking into
account both diagonal and nondiagonal coupling between different
exciton states~\cite{Muljarov04, Muljarov05}. In the present work,
we extend our theory to the nonlinear optical response of a
multilevel excitonic system coupled to acoustic phonons and
calculate the FWM polarization of the QD excited states.

\section{Optical response on a sequence of ultrashort pulses}

In the heterodyne technique used in~[1--\,4], semiconductor QDs
are excited by a sequence of ultrashort pulses:
\begin{equation}
{\cal E}(t)= E_1 \delta(t-t_1) +  E_2 \delta(t-t_2) + \dots,\ \ \
\ \ t_1< t_2< \dots
 \label{EF}
\end{equation}
The full Hamiltonian of the optically-driven exciton-phonon system
has the form ${\cal H}(t)=H+H_I(t)$, where $H$ is the
exciton-phonon Hamiltonian and $ H_I(t)=-d^\dagger {\cal E}(t) -
d\, {\cal E}^\ast(t) $  is the exciton-light interaction. Here
$d=\sum_{n} \mu_n \,|0\rangle\langle n|$ is the exciton dipole
moment operator, $|0\rangle$ is the QD vacuum state, and
$\mu_n\!=\!\int d{\bf r}\, \Psi_n({\bf r}_e\!=\!{\bf r},{\bf
r}_h\!=\!{\bf r})$ is the dipole moment of the given
single-exciton state $|n\rangle$ \,($n=1,\,2,\dots$).

The optical polarization then takes the form
\begin{equation}
{\cal P}(t;t_1,t_2,\dots)=\bigl\langle
U^\dagger(t)\,d(t)\,U(t)\bigr\rangle=\bigl\langle
U^\dagger_{E_1}(t_1) U^\dagger_{E_2}(t_2) \dots d(t) \dots
U_{E_2}(t_2) U_{E_1}(t_1)\bigr\rangle\,,
\end{equation}
where $d(t)= e^{iHt}\, d\,e^{-iHt}$. The particular form of the
electric field, Eq.~(\ref{EF}), allowed us to write the full
evolution operator $U(t)=T\exp\{-i\int_0^t e^{iHt'} H_I(t')
e^{-iHt'} dt'\}$ as a time-ordered product of operators $U_E(t)$
due to each individual pulse. The latter can be calculated
analytically in any order of $d$, giving up to first order
\begin{equation}
U_{E}(0)=e^{i(E\,d^\dagger+E^\ast d)}=\cos|\mu E|+i\frac{\sin|\mu
E|}{|\mu E|} (E\,d^\dagger+E^\ast d) +\dots,
 \label{UE}
\end{equation}
where $|\mu|^2=\sum_{n} |\mu_n|^2$. In the heterodyne technique
with a double-pulse excitation, the component proportional to
$E_1^\ast E_2^2$ is essentially filtered out from the full
polarization:
\begin{equation}
{\cal P}(t;t_1=-\tau,t_2=0)  \to -i\,\frac{\sin |2\mu E_1|}{2|\mu
E_1|} \left[\frac{\sin |\mu E_2|}{|\mu E|}\right]^2 \!E^\ast_1
E_2^2 \,P_{\rm FWM}(t,\tau)\,,
\end{equation}
where $P_{\rm FWM}(t,\tau)=\bigl\langle d(-\tau) d^\dagger(0) d(t)
d^\dagger(0)\bigr\rangle$ is the FWM polarization, $\tau>0$ is the
delay time between the two pulses.

\section{Excitonic multilevel system coupled to acoustic
phonons}

Neglecting biexcitonic effects, the Hamiltonian of several
excitonic states in a QD linearly coupled to acoustic phonons
takes the form~\cite{Muljarov05}
\begin{equation}
\hspace{-0mm} H=\sum_{n} \hbar\omega_n |n\rangle \langle
n|+\sum_{\bf q} \hbar\omega^{\rm ac}_{q}\, a^\dagger_{\bf q}
a_{\bf q}+\hbar V,\ \ \ \ \ V=\sum_{n,m}|n\rangle \langle
m|\sum_{\bf q} M_{nm}({\bf q} )\,  (a_{\bf q}+a^\dagger_{-\bf
q})\,,
\end{equation}
where $\hbar\omega_n $ is the bare exciton transition energy, and
$M_{nm}({\bf q} )$ the exciton-phonon matrix element. The FWM
polarization (defined in Sec.\,2) can be written as a product of
two infinite perturbation series
\begin{equation}
\hspace{-3mm}P_{\rm FWM}(t,\tau)=\!\sum_{n,m} |\mu_n|^2 |\mu_m|^2
e^{i(\omega_n \tau-\omega_m t)} \left\langle \langle n|{\cal
T}_{\rm inv}\, e^{\, i\!\int\limits_{-\tau}^{\,0} dt'\, V(t')}
|n\rangle\, \langle m|{\cal T}\,e^{ -i\int\limits_0^{\,t} dt''\,
V(t'')} |m\rangle \right\rangle,
 \label{series}
\end{equation}
where ${\cal T}_{\rm inv}$ arranges all negative-time operators in
the inverse order. The external brackets $\langle \dots \rangle$
in Eq.~(\ref{series}) mean the finite-temperature expectation
value which is taken over the phonon system and thus mixes
interaction operators from both series.
Following~\cite{Muljarov05} we use the cumulant expansion: For
each pair of exciton states $(n,m)$ we calculate the cumulant
$K(t,\tau)$ defined as
\begin{equation}
\bigl\langle \langle n|\dots |n\rangle \langle m|\dots |m\rangle
\bigr\rangle=1+P^{(1)}+P^{(2)}+\dots=\exp\bigl\{
{K^{(1)}+K^{(2)}+\dots} \bigr\}
\end{equation}
where $K^{(1)}=P^{(1)}$, \ $K^{(2)}=P^{(2)}-[ P^{(1)}]^2/2$, and
so on. In numerical calculations we proceed up to second order in
the cumulant thus taking into account both real and virtual
phonon-assisted transitions between different exciton
states~\cite{Muljarov05}.

In the special case of a \emph{single level}, the two series in
Eq.~(\ref{series}) can be combined into one, extending the time
integration from $-\tau$ to $t$, the interaction and time-order
operators being redefined, respectively, to
$\bar{V}(t)=\theta(t)V(t)$ and $\bar{{\cal T}}={\cal T}_{\rm
inv}$~$({\cal T})$ for negative (positive) times. As in the linear
polarization (and in any higher-order nonlinear response), the
cumulant expansion ends already in first order, giving exactly
\begin{equation}
P_{\rm FWM}(t,\tau)= e^{-i\omega_{\rm ex}
(t-\tau)}\exp\left\{-\frac{1}{2}\int_{\!-\tau}^t \!dt'
\int_{-\tau}^t \!dt'' \bigl\langle \bar{{\cal T}} \bar{V}(t')
\bar{V}(t'') \bigr\rangle\right\}=\frac{|P_L(t)|^2 \bigl[
P_L^\ast(\tau)\bigr]^2}{P_L^\ast(t+\tau)},
 \label{BM}
\end{equation}
where $P_L(t)=e^{-i\omega_{\rm ex} t}
\exp\bigl\{-\frac{1}{2}\int_0^t dt' \int_0^t dt'' \bigl\langle
{\cal T} V(t') V(t'') \bigr\rangle \bigr\}$ is the linear
polarization~\cite{Mahan}. Equation~(\ref{BM}) reproduces the
result by Vagov {\em et al.}~\cite{Vagov02}, but is derived here
in a more straightforward way.

\begin{figure}[htb]
\begin{minipage}[t]{.45\textwidth}
\includegraphics[angle=-90,width=\textwidth]{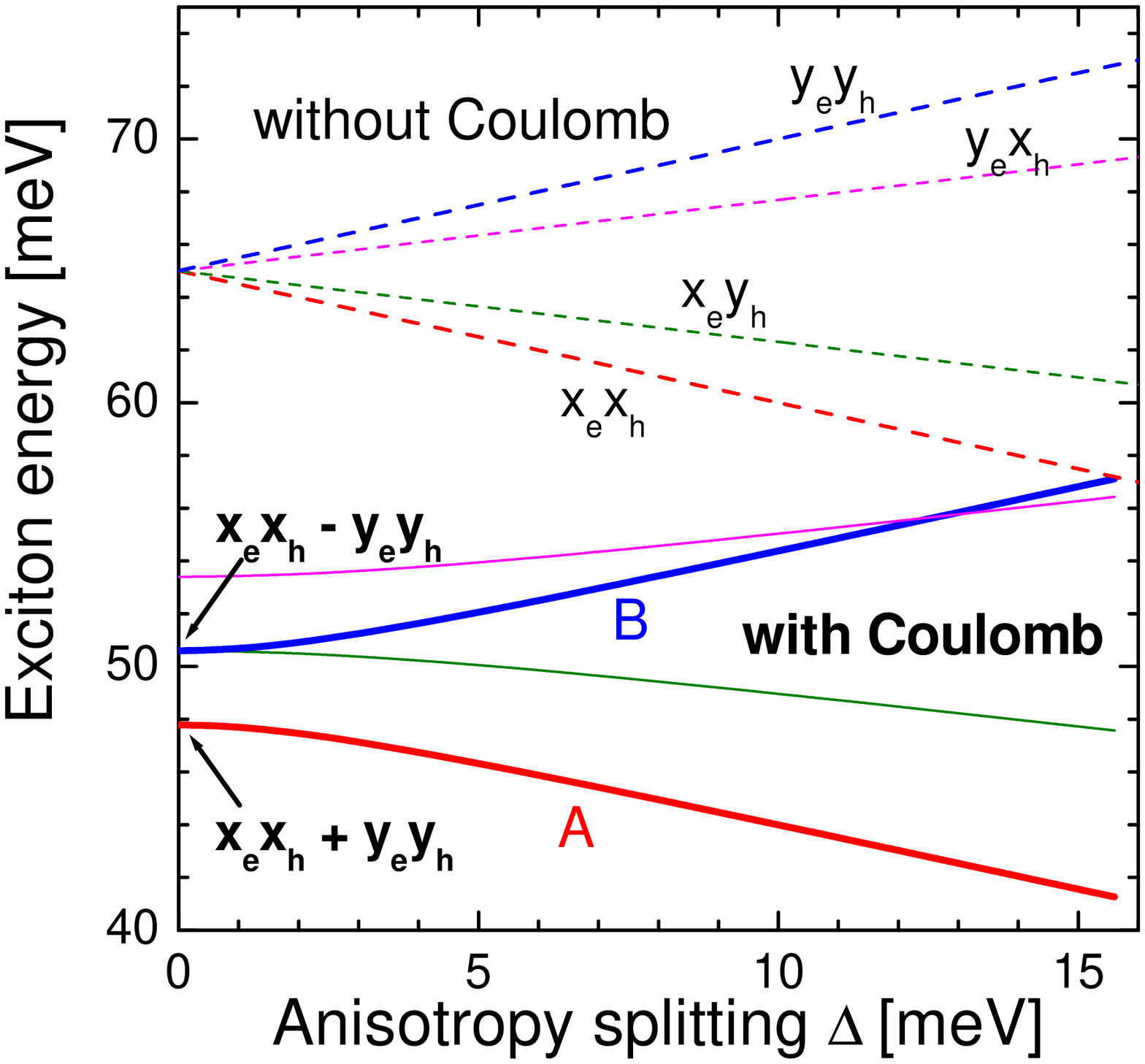}
\caption{Energies of exciton excited states in anisotropic pancake
InGaAs QDs calculated with (full curves) and without (dashed
curves) Coulomb interaction. A and B are the two bright excited
states (thick lines).} \label{fig:1}
\end{minipage}
\hfil
\begin{minipage}[t]{.45\textwidth}
\includegraphics[angle=-90,width=\textwidth]{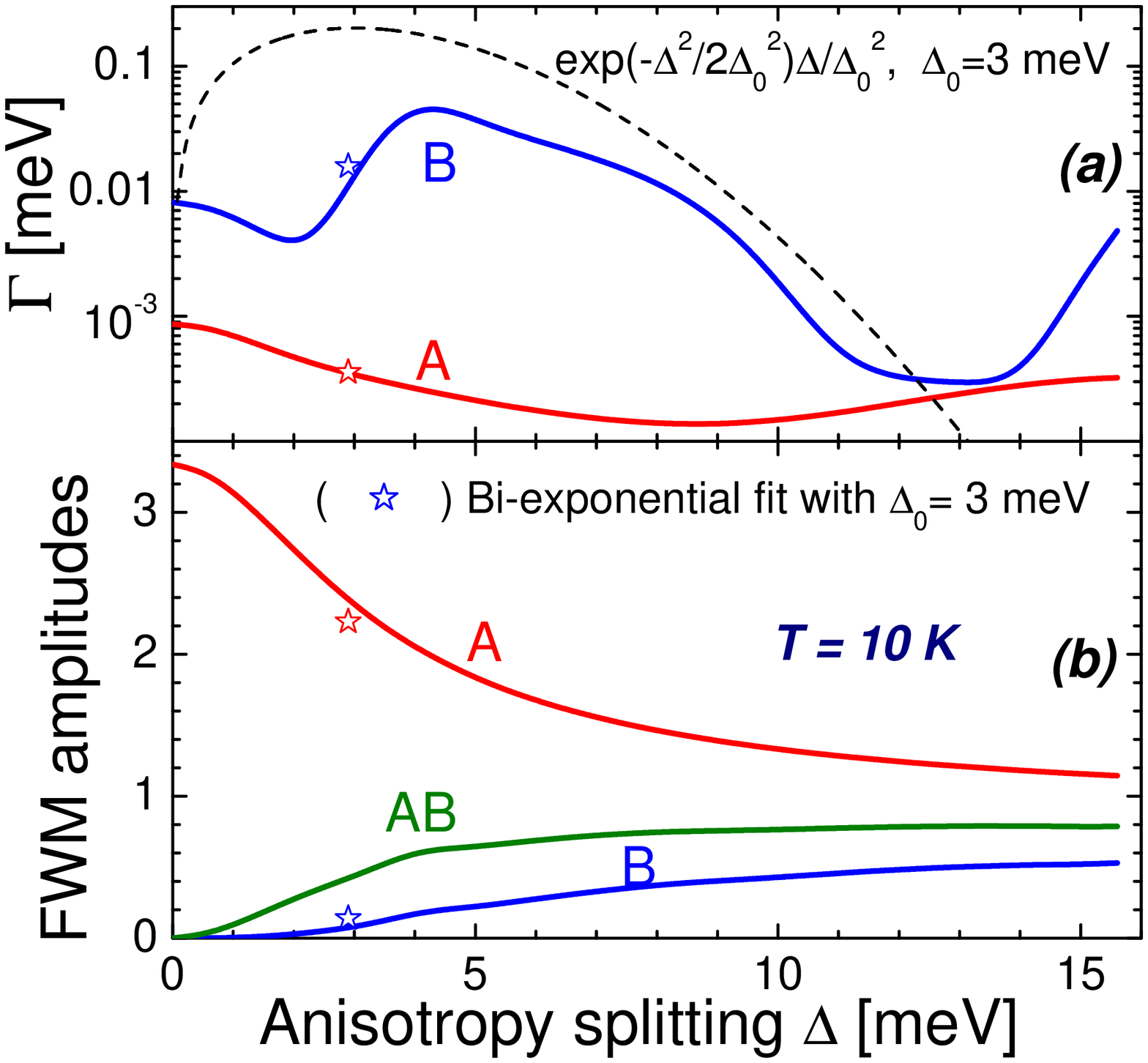}
\caption{Dephasing rates (a) and FWM amplitudes (b) of the bright
excited states at $T=10$\,K. Bi-exponential fit of the
ensemble-averaged photon echo is shown by stars. The splitting
distribution function is shown by a dashed line.} \label{fig:2}
\end{minipage}
\end{figure}

\section{Dephasing of QD excited states}

In order to simulate the measurable time-integrated FWM, we
consider an ensemble of QDs of different size and shape.
Variations of the QD size are mainly responsible for a wide
distribution ($\sim30$\,meV in InGaAs QDs~\cite{Borri05OECS}) of
the exciton transition frequency. Due to the phase prefactor in
Eq.~(\ref{series}), the transition frequency distribution leads to
the photon echo effect: The FWM polarization from different dots
destructively interfere at all times except those around $t=\tau$,
and the measured signal is $P_{\rm echo}(\tau)=P_{\rm
FWM}(\tau,\tau)$. In a proper simulation, the latter has to be
additionally averaged over a distribution of the QD excited states
splitting, which is much narrower
($\sim3$\,meV~\cite{Borri05OECS}) and is mainly due to QD shape
variations.

To do this, we assume a pancake shape of a QD (having a smaller
size in $z$-direction) with parabolic potentials for electrons and
holes, which are additionally anisotropic in the $xy$-plane. Here
we concentrate on exciton excited states formed from $x$- and
$y$-type electron and hole states. Without Coulomb interaction the
electron-hole pair has two bright ($x_ex_h$ and $y_ey_h$) and two
dark states ($x_ey_h$ and $y_ex_h$). Their energies are shown in
Fig.\,1 (dashed curves) as functions of the anisotropy splitting
$\Delta\equiv E_{y_ey_h}\!-\!E_{x_ex_h}$. Switching on the Coulomb
interaction adds to the exciton Hamiltonian a block-diagonal
matrix, the bright and the dark states not talking to each other
(due to parity), and brings in an additional energy splitting
within each doublet (Fig.\,1, solid curves).  At $\Delta=0$ bright
exciton states $A$ and $B$ are, respectively, exact symmetric and
antisymmetric combinations of the former states, the lower state
$A$ accumulating the whole oscillator strength [see Fig.\,2\,(b)].
\begin{SCfigure}[4][t]
\includegraphics[angle=0,width=.5\textwidth]{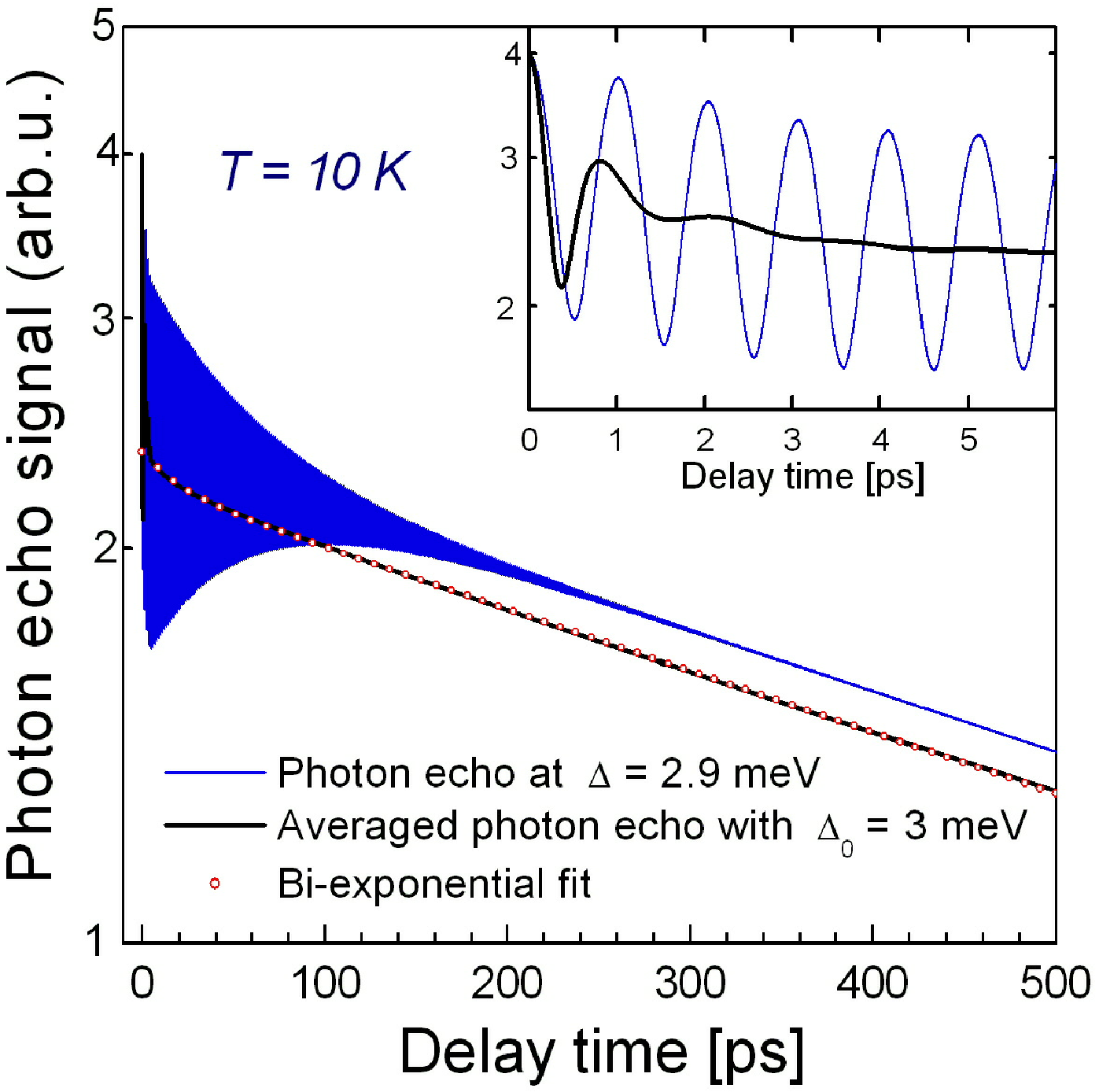}%
\caption{Photon echo signal of anisotropic QD calculated for
$\Delta=2.9$\,meV (thin curve) and the ensemble-averaged photon
echo (thick curve), at long times fitted by two exponents
(circles). Inset: the same in the short-time scale.} \label{fig:3}
\end{SCfigure}

Quite explicitly, for these excited states the photon echo can be
now written as
\begin{equation}
\hspace{-5mm} P_{\rm echo}(\tau)= \mu_A^4
\,C_A(\tau)\,e^{-2\Gamma_A\tau}+ \mu_B^4
\,C_B(\tau)\,e^{-2\Gamma_B\tau} + 2\mu_A^2\mu_B^2
\,C_{AB}(\tau)\,e^{-(\Gamma_A+\Gamma_B)\tau}\,\cos(\omega_A-\omega_B)\tau.
\label{echoAB}
\end{equation}
Remarkably, the dephasing rates $\Gamma_A$ and $\Gamma_B$ which
show up in the FWM polarization, Eq.~(\ref{echoAB}), turn out to
be exactly the same as in the linear polarization. The lower state
($A$) which is considerably split off has a rather weak dephasing
at $T=10$\,K [Fig.\,2\,(a)] which is mainly due to virtual
transitions. The upper state, in turn, is always close to one of
the dark states and has a much stronger dephasing due to real
transitions. It rapidly changes with $\Delta$ having maxima when
the level distance is close to the typical energy ($\sim2$\,meV)
of the acoustic phonons coupled to the QD. The photon echo
amplitudes  shown in Fig.\,2\,(b) are mainly determined by the
exciton dipole moments $\mu_A$, $\mu_B$. The functions $C_i(\tau)$
in Eq.~(\ref{echoAB}) change rapidly at short times, while
becoming constants in the long-time limit. These constant final
values are nothing else than the zero-phonon weights which at low
temperatures do not differ too much from unity.

The photon echo signal, Eq.~(\ref{echoAB}), shown in Fig.\,3 for
$\Delta=2.9$\,meV (thin curves) have oscillations up to 200\,ps
due to the mixed $AB$-term. The rapid short-time decay
(corresponding to the broadband in the spectrum) is also seen as
lowering of both maxima and minima (see the inset in Fig.\,3). We
finally perform the averaging over an ensemble of QDs having
different $xy$-anisotropy degree: $ \bar{P}_{\rm echo} (\tau)
=\int_0^\infty \!d\Delta\, P_{\rm
echo}(\tau;\Delta)\,e^{-\Delta^2/2\Delta_0^2}\,
\Delta/\Delta_0^{2} $, where $\Delta_0=3$\,meV is used. The
averaged function $ \bar{P}_{\rm echo} (\tau)$ has oscillations
only in the time scale of $2\pi\hbar/\Delta_0$ (Fig.\,3, thick
curves). At later times it is well approximated by a sum of two
exponents (Fig.\,3, circles) with decay times 20.8\,ps and
922\,ps. The amplitudes and half decay rates of the bi-exponential
fit are marked in Fig.\,2 by stars. Both amplitudes and rates are
very close to those of the photon echo signal ${P}_{\rm echo}
(\tau)$ of the particular QD with $\Delta=2.9$\,meV. Thus, we
believe that the measured double-exponential long-time
decay~\cite{Borri05OECS} is due to the two bright excited states
in QD, which have quite different dephasing times at low
temperatures.

\begin{acknowledgement}
We thank P. Borri and W. Langbein for stimulating discussions.
Financial support by DFG Sonderforschungsbereich 296 is gratefully
acknowledged. E.\,A.\,M. acknowledges partial support by Russian
Foundation for Basic Research and Russian Academy of Sciences.
\end{acknowledgement}


\begin{thebibliography}{10}
\bibitem{Borri05OECS} P. Borri and W. Langbein, 9th Conference on Optics and Excitons in Confined Systems,
Abstracts, Southampton (2005), p. 50.
\bibitem{Borri01} P. Borri, W. Langbein, S. Schneider, U. Woggon, R.\,L. Sellin,
D. Ouyang, and D. Bimberg, Phys. Rev. Lett. {\bf 87}, 157401
(2001).
\bibitem{Borri05} P. Borri, W. Langbein, U. Woggon, V. Stavarache, D. Reuter, and A.\,D. Wieck,
Phys. Rev. B {\bf 71}, 115328 (2005).
\bibitem{Borri03} P. Borri, W. Langbein, U. Woggon, M. Schwab, M. Bayer, S. Fafard,
Z. Wasilewski, and P. Hawrylak, Phys. Rev. Lett. {\bf 91}, 267401
(2003).
\bibitem{Vagov02} A. Vagov, V.\,M. Axt, and T. Kuhn,  Phys. Rev.
B {\bf 66}, 165312 (2002).
\bibitem{Vagov03} A. Vagov, V.\,M. Axt, and T. Kuhn, Phys. Rev.
B {\bf 67}, 115338 (2003).
\bibitem{Mahan} G. Mahan, Many-Particle Physics, (Plenum, New York, 1990).
\bibitem{Vagov04} A. Vagov, V.\,M. Axt, and T. Kuhn, W. Langbein, P. Borri, and U. Woggon,
Phys. Rev. B {\bf 70}, 201305 (2004).
\bibitem{Muljarov04} E.\,A. Muljarov and R. Zimmermann, Phys. Rev. Lett. {\bf 93}, 237401 (2004).
\bibitem{Muljarov05} E.\,A. Muljarov, T. Takagahara, and R. Zimmermann,
Phys. Rev. Lett. {\bf 95}, 177405 (2005).

\end{thebibliography}
\end{document}